\documentclass[twocolumn,preprintnumbers,amsmath,amssymb]{revtex4-1}

\usepackage{amstext,amssymb,amsfonts}
\usepackage{bm}
\usepackage{amsmath}

\usepackage{graphicx}
\usepackage{graphics}
\usepackage{epsfig}

\usepackage{txfonts}

\usepackage{hyperref}

\newcommand{\bra}[1]{\ensuremath{\langle #1 \vert}}
\newcommand{\ket}[1]{\ensuremath{\vert #1  \rangle}}

\renewcommand{\b}[1]{\ensuremath{\mathbf{#1}}}
\renewcommand{\H}{\ensuremath{\text{H}}}
\renewcommand{\l}{\ensuremath{\lambda}}

\renewcommand{\d}{\ensuremath{\text{d}}}

\DeclareMathOperator{\erf}{erf}

\newcommand{\isEquivTo}[1]{\underset{#1}{\sim}}

\begin{document}
\title{Short-range correlation energy of the relativistic homogeneous electron gas}
\author{Julien Paquier$^1$} \email{julien.paquier@lct.jussieu.fr}
\author{Julien Toulouse$^{1,2}$} \email{toulouse@lct.jussieu.fr}
\affiliation{$^1$Laboratoire de Chimie Th\'eorique (LCT), Sorbonne Universit\'e and CNRS, F-75005 Paris, France}
\affiliation{$^2$Institut Universitaire de France, F-75005 Paris, France}
\date{March 31, 2021}

\begin{abstract}
We construct the complementary short-range correlation relativistic local-density-approximation functional to be used in relativistic range-separated density-functional theory based on a Dirac-Coulomb Hamiltonian in the no-pair approximation. For this, we perform relativistic random-phase-approximation calculations of the correlation energy of the relativistic homogeneous electron gas with a modified electron-electron interaction, we study the high-density behavior, and fit the results to a parametrized expression. The obtained functional should eventually be useful for electronic-structure calculations of strongly correlated systems containing heavy elements.
\end{abstract}

\maketitle

\section{introduction}

Range-separated density-functional theory (RS-DFT) (see, e.g., Refs.~\onlinecite{Sav-INC-96,TouColSav-PRA-04}) is an extension of Kohn-Sham density-functional theory (KS-DFT)~\cite{KohSha-PR-65} which allows one to rigorously combine a multideterminant wave-function method accounting for the long-range part of the electron-electron interaction with a complementary short-range density functional. RS-DFT can improve over usual Kohn-Sham density-functional approximations for the electronic-structure calculations of strongly correlated systems (see, e.g., Refs.~\onlinecite{HedTouJen-JCP-18,FerGinTou-JCP-19}) and/or systems involving weak intermolecular interactions (see, e.g., Refs.~\onlinecite{TayAngGalZhaGygHirSonRahLilPodBulHenScuTouPevTruSza-JCP-16,KalMusTou-JCP-19}), while still enjoying a fast basis convergence~\cite{FraMusLupTou-JCP-15}.

With the aim of describing compounds with heavy elements which involve both strong correlation and relativistic effects, RS-DFT has been extended to a four-component relativistic framework~\cite{KulSau-CP-12,SheKneSau-PCCP-15,PaqTou-JCP-18,PaqGinTou-JCP-20}. In this relativistic RS-DFT, the no-pair~\cite{Suc-PRA-80,Mit-PRA-81} ground-state electronic energy of the Dirac-Coulomb Hamiltonian is written as~\cite{PaqGinTou-JCP-20} 
\begin{eqnarray}
E_0  &=& \bra{\Psi_+} \hat{T}_\text{D} + \hat{V}_\text{ne} + \hat{W}_{\text{ee}}^{\text{lr},\mu} \ket{\Psi_+} + \bar{E}_\text{Hxc}^{\text{sr},\mu}[n_{\Psi_+}],\;\;\;\;\;
\label{ERSDFT}
\end{eqnarray}
where $\hat{T}_\text{D}$ is the kinetic + rest mass Dirac operator, $\hat{V}_{\text{ne}}$ is the nuclei-electron interaction operator, $\hat{W}_{\text{ee}}^{\text{lr},\mu}$ is the electron-electron interaction operator associated with the long-range pair potential $w_{\text{ee}}^{\text{lr},\mu}(r_{12}) = \erf(\mu r_{12})/r_{12}$, and $\bar{E}_\text{Hxc}^{\text{sr},\mu}[n_{\Psi_+}]$ is the corresponding complementary short-range relativistic Hartree-exchange-correlation functional evaluated at the density of $\Psi_+$. The no-pair multideterminant wave function $\Psi_+$ is constructed from positive-energy states only and can in principle be obtained using a minmax principle~\cite{Tal-PRL-86,DatDev-PJP-88,GriSie-JLMS-99,DolEstSer-JFA-00,SauVis-INC-03,AlmKneJenDyaSau-JCP-16,PaqTou-JCP-18,PaqGinTou-JCP-20}. The range-separation parameter $\mu \in [0,+\infty)$ controls the range of the separation. For $\mu=0$, the long-range interaction vanishes and no-pair relativistic KS-DFT (see, e.g., Refs.~\onlinecite{Eng-INC-02,SauHel-JCC-02}) is recovered. For $\mu\to\infty$, the long-range interaction reduces to the full-range Coulomb interaction and no-pair relativistic wave-function theory (see, e.g., Refs.~\onlinecite{SauVis-INC-03,Liu-JCP-20}) is recovered.

While any existing wave-function approximation can directly be used for ${\Psi_+}$, new approximations need to be developed for the short-range relativistic functional $\bar{E}_\text{Hxc}^{\text{sr},\mu}[n]$. As usual, this functional can be decomposed into Hartree, exchange, and correlation contributions
\begin{eqnarray}
\bar{E}_\text{Hxc}^{\text{sr},\mu}[n] = E_\text{H}^{\text{sr},\mu}[n] + E_\text{x}^{\text{sr},\mu}[n] + \bar{E}_\text{c}^{\text{sr},\mu}[n].
\label{EHxcsrmu}
\end{eqnarray}
The short-range Hartree functional is
\begin{eqnarray}
E_\text{H}^{\text{sr},{\mu}}[n] = \frac{1}{2} \iint n(\b{r}_1) n(\b{r}_2) w_{\text{ee}}^{\text{sr},\mu}(r_{12}) \d\b{r}_1 \d\b{r}_2,
\label{EHsrmu}
\end{eqnarray}
where $w_{\text{ee}}^{\text{sr},\mu}(r_{12})=1/r_{12} - w_{\text{ee}}^{\text{lr},\mu}(r_{12})$ is the short-range pair potential. The short-range exchange functional is
\begin{eqnarray}
E_{\text{x}}^{\text{sr},{\mu}}[n] = \bra{\Phi_+[n]} \; \hat{W}_\text{ee}^{\text{sr},{\mu}} \; \ket{\Phi_+[n]} - E_\text{H}^{\text{sr},{\mu}}[n],
\label{Exsrmu}
\end{eqnarray}
where $\Phi_+[n]$ is the relativistic Kohn-Sham single-determinant wave function and $\hat{W}_\text{ee}^{\text{sr},{\mu}}$ is the electron-electron interaction operator associated with $w_{\text{ee}}^{\text{sr},\mu}(r_{12})$. In Refs.~\onlinecite{KulSau-CP-12,SheKneSau-PCCP-15}, the relativistic short-range exchange and correlation functionals $E_{\text{x}}^{\text{sr},{\mu}}[n]$ and $\bar{E}_\text{c}^{\text{sr},\mu}[n]$ were approximated by non-relativistic short-range exchange and correlation functionals, which is a reasonable first approximation since for valence properties relativistic effects are usually dominated by the kinematic contribution and the induced change in the density (see, e.g., Ref.~\onlinecite{Wul-INC-10}). To go beyond this non-relativistic approximation and put relativistic RS-DFT on a firmer ground, we have constructed for the short-range exchange functional $E_{\text{x}}^{\text{sr},{\mu}}[n]$ the relativistic local-density approximation (RLDA) in Ref.~\onlinecite{PaqTou-JCP-18} and approximations going beyond the RLDA in Ref.~\onlinecite{PaqGinTou-JCP-20}. In the present work, we turn to the short-range correlation functional $\bar{E}_\text{c}^{\text{sr},\mu}[n]$ and we develop the RLDA for it.

The complementary short-range correlation RLDA functional is defined as
\begin{eqnarray}
\bar{E}_{\text{c}}^{\text{sr,RLDA},{\mu}}[n] &=& \int \text{d}\b{r}~n(\b{r})~\bar{\epsilon}_{\text{c}}^{\text{sr,RHEG},\mu}(n(\b{r})),
\end{eqnarray} 
with
\begin{eqnarray}
\bar{\epsilon}_{\text{c}}^{\text{sr,RHEG},\mu}(n) = {\epsilon}_{\text{c}}^{\text{RHEG}}(n) - \epsilon_{\text{c}}^{\text{lr,RHEG},\mu}(n),
\label{epscsrRHEG}
\end{eqnarray} 
where ${\epsilon}_{\text{c}}^{\text{RHEG}}(n)$ and $\epsilon_{\text{c}}^{\text{lr,RHEG},\mu}(n)$ are the correlation energies per particle of the relativistic homogeneous electron gas (RHEG) with full-range and long-range electron-electron interactions, respectively. We express each of these correlation energies per particle as the correlation energy per particle of the corresponding non-relativistic homogeneous electron gas (HEG) multiplied by a relativistic correlation factor 
\begin{eqnarray}
	{\epsilon}_{\text{c}}^{\text{RHEG}}(n) = {\epsilon}_{\text{c}}^{\text{HEG}}(n)~{\phi}_{\text{c}}(n),
	\label{eq:full_range_correlation_density_functional}
\end{eqnarray} 
\begin{eqnarray}
{\epsilon}_{\text{c}}^{\text{lr,RHEG},\mu}(n) = {\epsilon}_{\text{c}}^{\text{lr,HEG},\mu}(n)~{\phi}_{\text{c}}^{\tilde{\mu}}(n),
        \label{eq:long_range_correlation_density_functional}
\end{eqnarray}
where we have introduced the scaled range-separation parameter 
\begin{eqnarray}
\tilde{\mu} = \frac{\mu}{k_{\text{F}}},
\end{eqnarray}
where $k_{\text{F}} = (3{\pi}^{2}n)^{1/3}$ is the Fermi wave vector. The scaled range-separation parameter $\tilde{\mu}$ is a natural adimensional parameter measuring the range of the interaction relative to the density. We must have ${\phi}_{\text{c}}^{\text{lr},\tilde{\mu}\to\infty}(n) = {\phi}_{\text{c}}(n)$ since the long-range interaction reduces to the full-range one in this limit. Equations~(\ref{eq:full_range_correlation_density_functional}) and~(\ref{eq:long_range_correlation_density_functional}) allow one to use already existing parametrizations for ${\epsilon}_{\text{c}}^{\text{HEG}}(n)$ and ${\epsilon}_{\text{c}}^{\text{lr,HEG},{\mu}}(n)$~\cite{PerWan2-PRB-92,PazMorGorBac-PRB-06}. 

The correlation energy per particle of the RHEG ${\epsilon}_{\text{c}}^{\text{RHEG}}(n)$ was first estimated at the random-phase approximation (RPA) level by Ramana and Rajagopal~\cite{RamRaj-PRA-81} (see also Refs.~\onlinecite{EngMulSpeDre-INC-95,EngKelFacMulDre-PRA-95,EngDre-INC-96,FacEngDreAndMul-PRA-98,Eng-INC-02}), and the corresponding relativistic correlation factor ${\phi}_{\text{c}}(n)$ was parametrized by Schmid \textit{et al.}~\cite{SchEngDreBlaSch-AQC-98}. In the same spirit, we estimate in this work the relativistic long-range correlation factor ${\phi}_{\text{c}}^{\text{lr},\tilde{\mu}}(n)$ at the RPA level, i.e.
\begin{eqnarray}
{\phi}_{\text{c}}^{\text{lr},\tilde{\mu}}(n) \approx {\phi}_{\text{c}}^{\text{lr,RRPA},\tilde{\mu}}(n) = \frac{{\epsilon}_{\text{c}}^{\text{lr,RRPA},\tilde{\mu}}(n)}{{\epsilon}_{\text{c}}^{\text{lr,RPA},\tilde{\mu}}(n)},
\end{eqnarray}
where ${\epsilon}_{\text{c}}^{\text{lr,RRPA},\tilde{\mu}}(n)$ is the long-range relativistic random-phase-approximation (RRPA) correlation energy per particle of the RHEG and ${\epsilon}_{\text{c}}^{\text{lr,RPA},\tilde{\mu}}(n)$ is its non-relativistic analog. The use of the RPA appears consistent considering that relativistic effects are most important in the high-density regime, for which the RPA provides a good approximation to the correlation energy. Contrary to the RRPA calculations of Ramana and Rajagopal~\cite{RamRaj-PRA-81} which included the transverse contribution from the full quantum-electrodynamics (QED) photon propagator and were performed within the no-sea approximation (i.e., including a renormalization contribution from negative-energy states)~\cite{FacEngDreAndMul-PRA-98}, here our RRPA calculations are limited to the longitudinal component of the interaction in the Coulomb gauge and within the no-pair approximation. We do so for consistency since in the relativistic RS-DFT of Eq.~(\ref{ERSDFT}), the long-range wave-function part is treated at the same level. The numerically calculated relativistic long-range correlation factor ${\phi}_{\text{c}}^{\text{lr,RRPA},\tilde{\mu}}(n)$ is then fitted to a parametrized expression imposing the correct high-density limit.

Hartree atomic units (a.u.) are used throughout the paper.

\section{Long-range correlation energy from random-phase approximation}

\subsection{Relativistic random-phase approximation}

\begin{figure*}[t]
        \centering
        \includegraphics[width=5.9cm,angle=270]{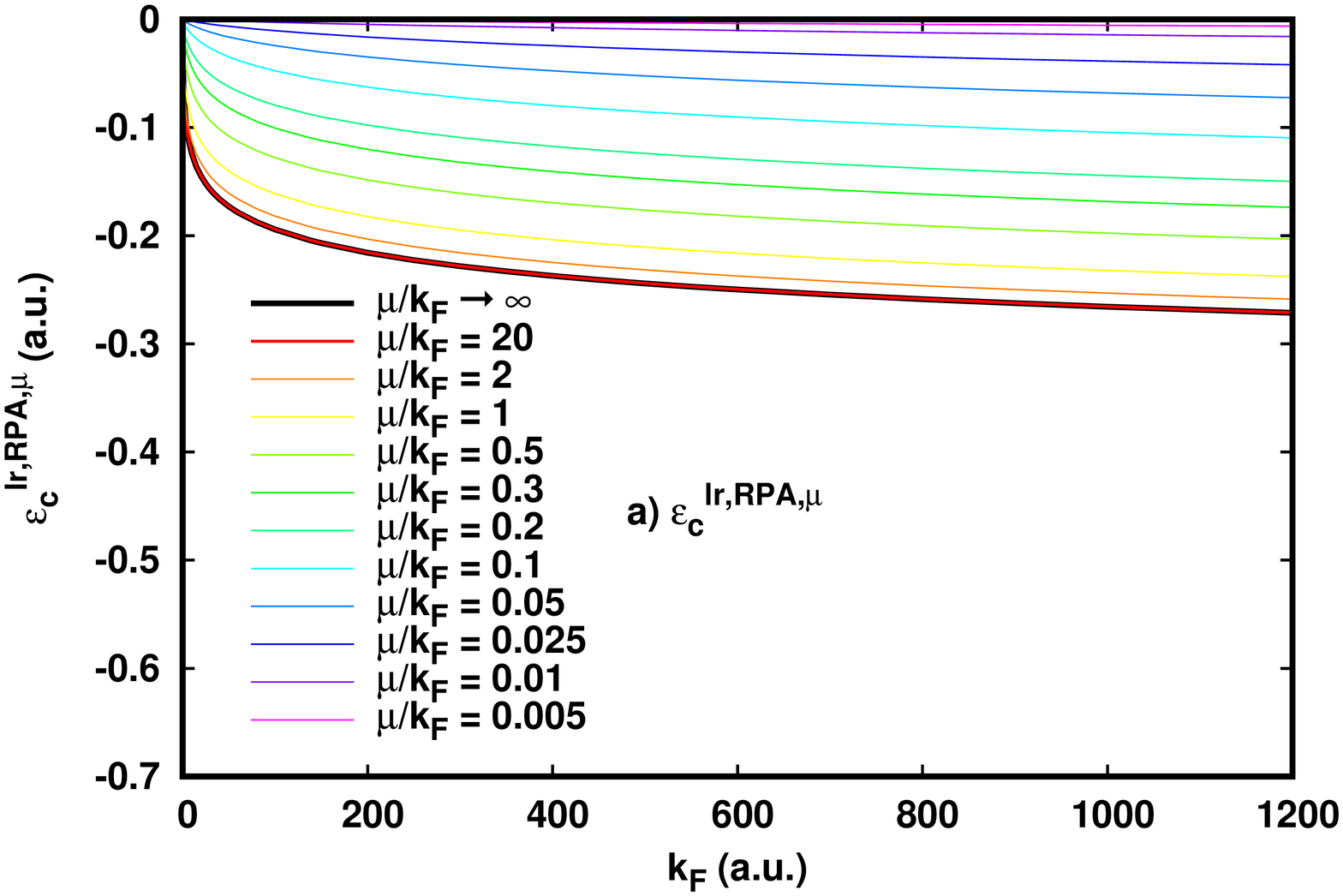}
        \includegraphics[width=5.9cm,angle=270]{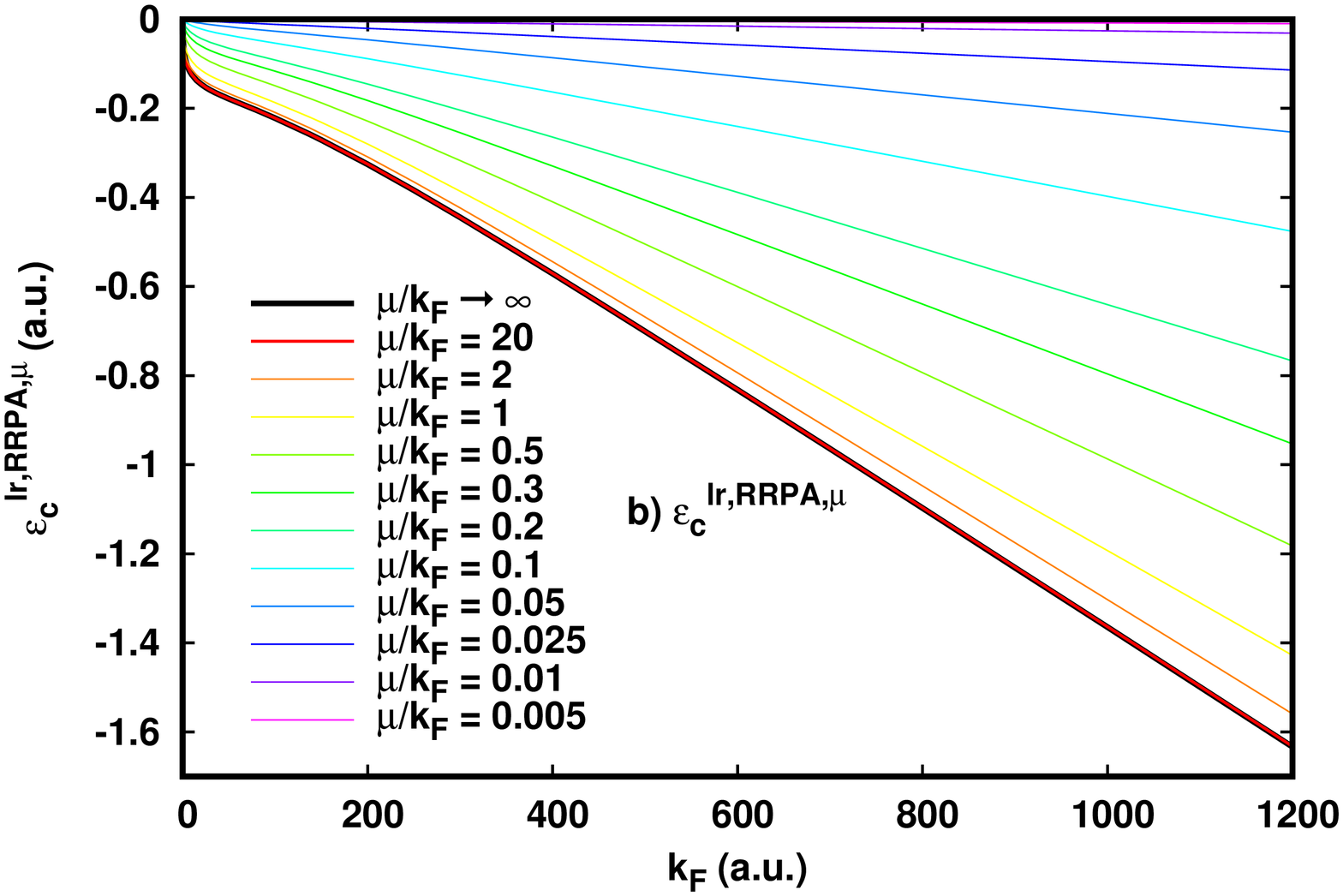}
\caption{Non-relativistic (a) and relativistic (b) long-range RPA correlation energies per particle of the HEG.}
\label{fig:correlation_energy}
\end{figure*}

As already indicated, we want to determine the long-range RRPA correlation energy per particle of the RHEG within the no-pair approximation and for the longitudinal component of the electron-electron interaction in the Coulomb gauge. With these approximations, the expression of ${\epsilon}_{\text{c}}^{\text{lr,RRPA},{\tilde{\mu}}} (n)$ is the same as its non-relativistic counterpart (see, e.g., Refs.~\onlinecite{LeiGroPer-PRB-00,Tou-PRB-05})
 \begin{widetext}
	 \begin{eqnarray}
	{\epsilon}_{\text{c}}^{\text{lr,RRPA},{\tilde{\mu}}} (n)= -\frac{1}{2{\pi}~n}\int \frac{\text{d}\b{q}}{(2{\pi})^{3}} w^{\text{lr},\tilde{\mu}}(q)\int_{0}^{\infty}\text{d}u  \int_{0}^{1}\text{d}{\lambda} \frac{\left[{\chi}_{0}(q,iu)\right]^{2} f_{\H}^{\text{lr},\tilde{\mu},{\lambda}}(q)}{1-{\chi}_{0}(q,iu)f_{\H}^{\text{lr},\tilde{\mu},{\lambda}}(q)},
	\label{eq:correlation_energy}
\end{eqnarray}
where $\l$ is a coupling constant. In this expression, ${\chi}_{0}(q,iu)$ is the relativistic longitudinal non-interacting linear-response function of the RHEG within the no-pair approximation at wave vector $q=|\b{q}|$ and imaginary frequency $iu$ (see Refs.~\onlinecite{Jan-NC-62,RamRaj-PRA-81,FacEngDreAndMul-PRA-98} and Appendix~\ref{app:linear_response_function})
	\begin{eqnarray}
	{\chi}_{0}(q,iu) &=& - k_{\text{f}} \int \frac{\text{d}\tilde{\b{k}}}{(2{\pi})^{3}} {\theta}(1 - \tilde{k})
        \frac{\bigg(\sqrt{\tilde{k}^{2} + \tilde{c}^{2}} + \sqrt{|\tilde{\b{k}}+\tilde{\b{q}}|^{2} + \tilde{c}^{2}} \bigg)^{2} - \tilde{q}^{2}}{\sqrt{\tilde{k}^{2} + \tilde{c}^{2}}\sqrt{|\tilde{\b{k}}+\tilde{\b{q}}|^{2} + \tilde{c}^{2}}} \frac{ \tilde{c} \bigg(\sqrt{|\tilde{\b{k}}+\tilde{\b{q}}|^{2} + \tilde{c}^{2}}-\sqrt{\tilde{k}^{2} + \tilde{c}^{2}}\bigg)}{ {\tilde{u}}^{2}+ \tilde{c}^{2}\bigg( \sqrt{|\tilde{\b{k}}+\tilde{\b{q}}|^{2} + \tilde{c}^{2}}-\sqrt{\tilde{k}^{2} + \tilde{c}^{2}} \bigg)^{2}},
	\label{eq:linear_response_fct}
\end{eqnarray}
 \end{widetext}
where we have introduced the adimensional variables 
\begin{eqnarray}
\tilde{\b{k}} = \frac{\b{k}}{k_{\text{F}}}, \; \; \tilde{\b{q}} = \frac{\b{q}}{k_{\text{F}}}, \;\; \tilde{u} = \frac{u}{k_{\text{F}}^{2}}, \;\; \tilde{c} = \frac{c}{k_{\text{F}}},
\end{eqnarray}
where $c= 137.036$ a.u. is the speed of light. Note that the scaled speed of light $\tilde{c}$ is a natural adimensional parameter measuring the importance of relativistic effects (relativistic effects are negligible for $\tilde{c} \gg 1$ and increase as $\tilde{c}$ decreases). In Eq.~(\ref{eq:correlation_energy}), $f_{\H}^{\text{lr},\tilde{\mu},{\lambda}}(q)$ is the long-range Hartree kernel at the coupling constant $\l$ given by the Fourier transform of the long-range interaction 
\begin{eqnarray}
	f_{\H}^{\text{lr},\tilde{\mu},{\lambda}}(q) ~~=~~ {\lambda}w^{\text{lr},\tilde{\mu}}(q) &=&  {\lambda}~\frac{4{\pi}}{\tilde{q}^{2}k_{\text{F}}^{2}}\text{exp}\left[\frac{-\tilde{q}^{2}}{4\tilde{\mu}^{2}} \right].
	\label{eq:Hartree_kernel}
\end{eqnarray}
Performing the integrals in Eq.~(\ref{eq:correlation_energy}) over the angular variables of $\b{q}$ and over the coupling constant ${\lambda}$ gives
\begin{widetext}
	\begin{eqnarray}
	{\epsilon}_{\text{c}}^{\text{lr,RRPA},{\tilde{\mu}}}(n) 
	&=&  -\frac{3}{4{\pi}} \int_{0}^{\infty} \text{d}\tilde{q} ~\int_{0}^{\infty}\text{d}{\tilde{u}} ~ \Bigg(4{\pi}~\text{exp}\left[\frac{-\tilde{q}^{2}}{4\tilde{\mu}^{2}} \right]{\chi}_{0}(\tilde{q}k_{\text{F}},i\tilde{u} k_{\text{F}}^2) 
		+ \tilde{q}^{2}k_{\text{F}}^{2}~\text{ln}\left[1 - \frac{4{\pi}}{\tilde{q}^{2}k_{\text{F}}^{2}}\text{exp}\left[\frac{-\tilde{q}^{2}}{4\tilde{\mu}^{2}} \right] {\chi}_{0}(\tilde{q}k_{\text{F}},i\tilde{u} k_{\text{F}}^2) \right]\Bigg). 
	\label{eq:correlation_energy_adimensional_variables}
\end{eqnarray}
\end{widetext}
As in the non-relativistic case, the integral over $\tilde{q}$ and $\tilde{u}$ are performed numerically. However, contrary to the non-relativistic case, we also do numerically the integral over $\tilde{\b{k}}$ in the linear-response function in Eq.~(\ref{eq:linear_response_fct}). In total, this gives a four-dimensional numerical integration that we calculate using the software Wolfram Mathematica~\cite{Math12-PROG-20} with six digits of accuracy. In the non-relativistic limit, i.e. $\tilde{c} \rightarrow \infty$, the integral defining the linear-response function in Eq.~(\ref{eq:linear_response_fct}) can easily be done analytically and Eq.~(\ref{eq:linear_response_fct}) reduces to the well-known non-relativistic Lindhard function~\cite{BarHed-JPC-72}. However, for consistency, we also use a four-dimensional numerical integration with the same precision for $\tilde{c} \rightarrow \infty$ to obtain the non-relativistic RPA long-range correlation energy per particle ${\epsilon}_{\text{c}}^{\text{lr,RPA},{\tilde{\mu}}}(n) = \lim_{\tilde{c} \rightarrow \infty} {\epsilon}_{\text{c}}^{\text{lr,RRPA},{\tilde{\mu}}}(n)$. We use 41 values of the Fermi wave vector $k_{\text{F}}$ ranging from 0.005 to 1200 a.u. (corresponding to a range of Wigner-Seitz radius $r_\text{s} = [3/(4\pi n)]^{1/3}$ from 384 to 0.0016 a.u.). The highest sampled density corresponds to more than twice the maximal core electronic density of uranium, thus encompassing all chemically relevant electronic densities. For the scaled range-separation parameter $\tilde{\mu} = {\mu}/k_{\text{F}}$, we consider 25 different values ranging from 0.005 to 20 a.u., in addition to the $\tilde{\mu} \rightarrow \infty$ limit giving the full-range RRPA and RPA correlation energies ${\epsilon}_{\text{c}}^{\text{lr,RRPA},{\tilde{\mu}\to\infty}}(n)={\epsilon}_{\text{c}}^{\text{RRPA}}(n)$ and ${\epsilon}_{\text{c}}^{\text{lr,RPA},{\tilde{\mu}\to\infty}}(n)={\epsilon}_{\text{c}}^{\text{RPA}}(n)$. Note that the speed of light $c$ is fixed to its physical value in our calculations, i.e. we do not try to obtain the dependence on $c$ of the RRPA correlation energy. For more details on the numerical calculations, see Ref.~\onlinecite{Paq-THESIS-20}.

\subsection{Long-range correlation energy}
\label{subsec:long_range_correlation_energy}

\begin{figure*}[t]
        \centering
         \includegraphics[width=5.9cm,angle=270]{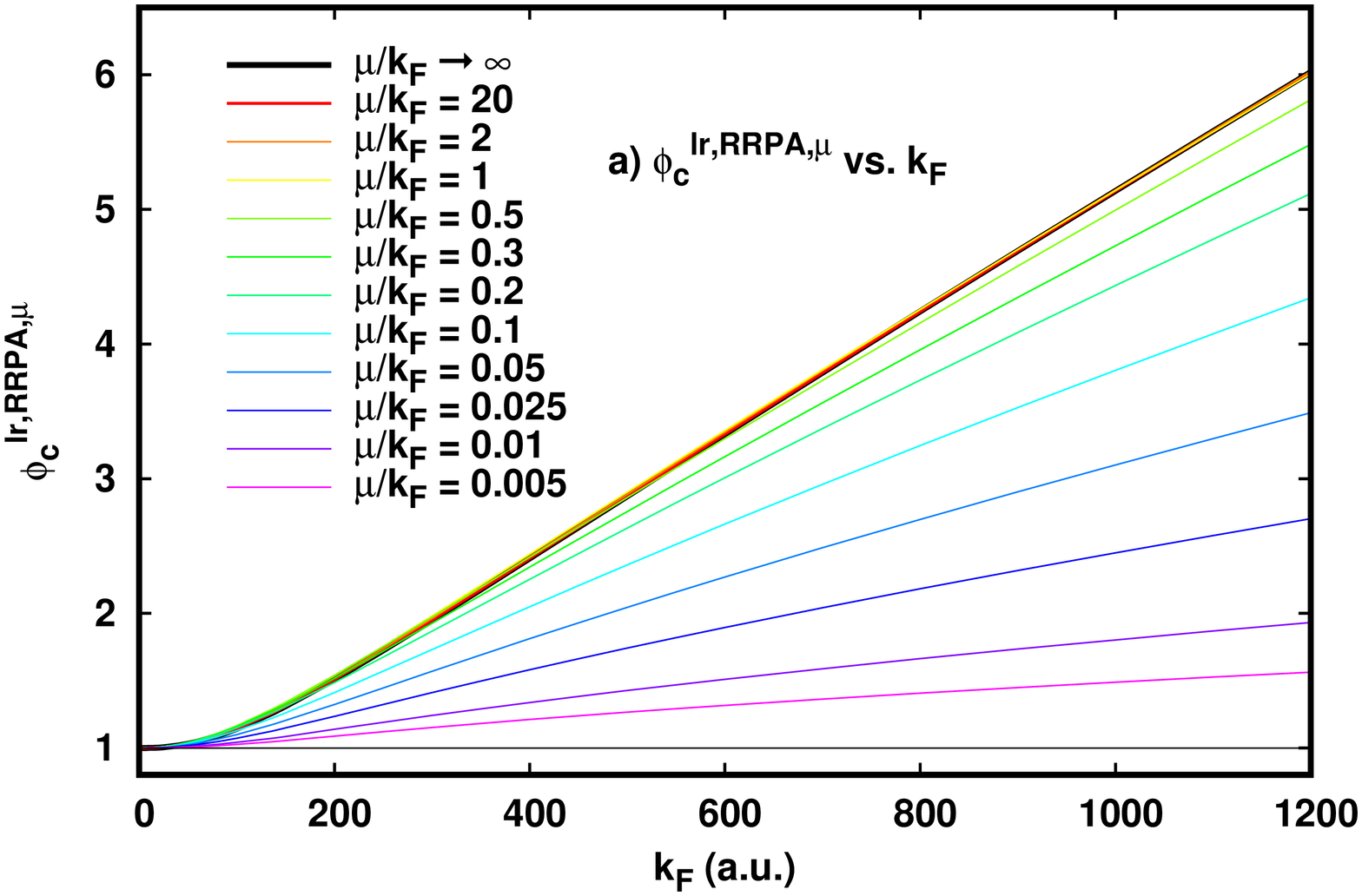}
         \includegraphics[width=5.9cm,angle=270]{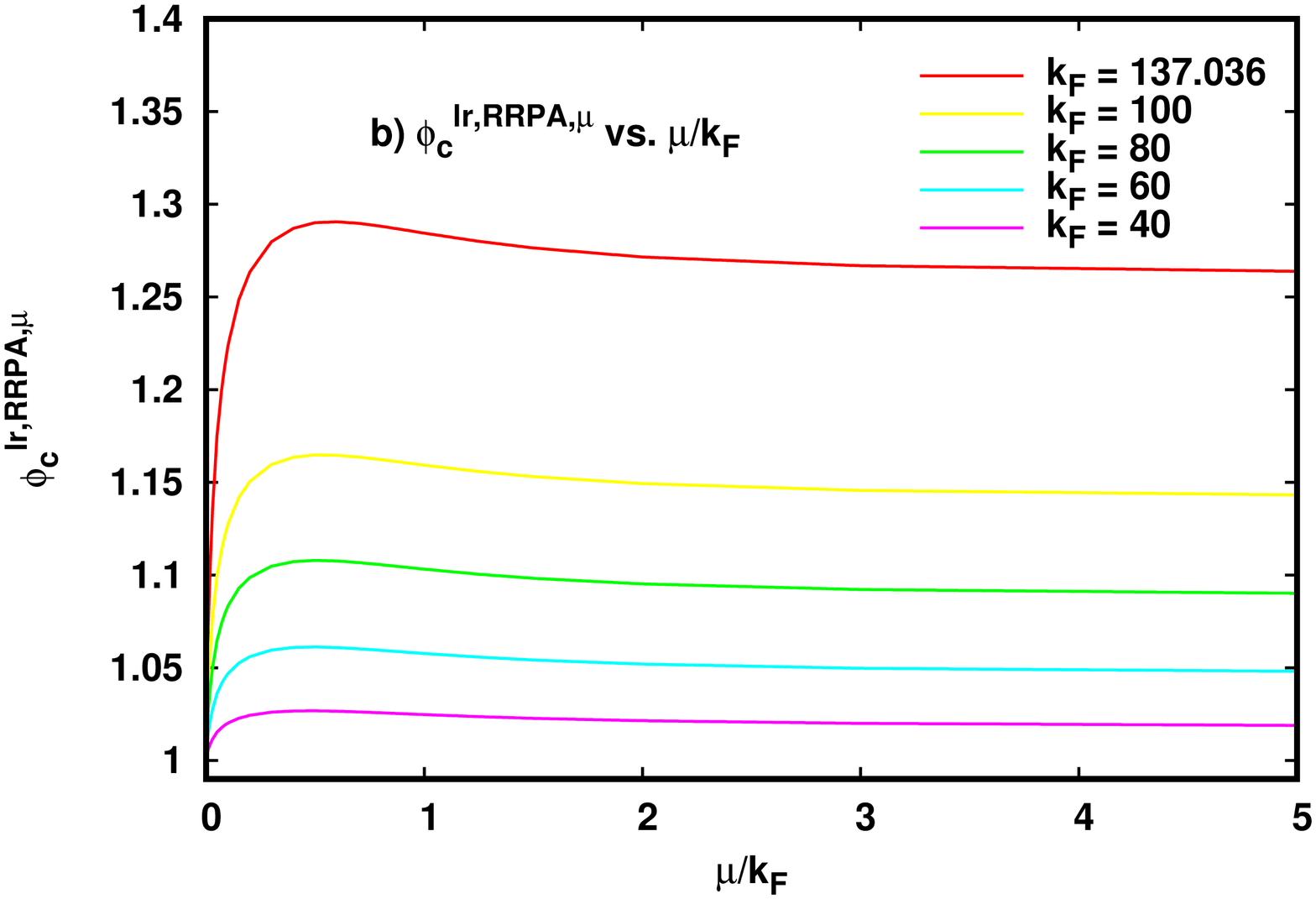}
	\caption{Relativistic long-range correlation factor ${\phi}_{\text{c}}^{\text{lr,RRPA},\tilde{\mu}}(n)$ as a function of $k_{\text{F}}$ (a) and $\tilde{\mu}=\mu/k_{\text{F}}$ (b).}
\label{fig:phi_lr}
\end{figure*}

We show in Fig.~\ref{fig:correlation_energy} the non-relativistic and relativistic long-range RPA correlation energies per particle as a function of $k_{\text{F}}$ for several values of $\tilde{\mu}$. As regards the non-relativistic results, for $\tilde{\mu} \to \infty$, we correctly reproduce the high-density expansion of the full-range RPA correlation energy per particle (see, e.g., Ref.~\onlinecite{PerWan-PRB-92}) that we expressed here in terms of $k_{\text{F}}$
\begin{eqnarray}
	{\epsilon}_{\text{c}}^{\text{RPA}}(n)
	&=& - \frac{1 - \ln 2}{{\pi}^{2}}\ln k_{\text{F}} -0.05083 + O\left(\frac{\ln k_{\text{F}}}{k_{\text{F}}}\right).
	\label{eq:full_range_high_density}
\end{eqnarray}
This is the usual weak-correlation limit where the correlation energy per particle is negligible compared to the exchange energy per particle which is linear in $k_{\text{F}}$. We observe a similar logarithmic behavior also for the long-range RPA correlation energy per particle on our chosen range of $k_{\text{F}}$ for values of $\tilde{\mu}$ larger than $0.1$ a.u.. For $\tilde{\mu} \gtrsim 20$ a.u., the long-range RPA correlation energy is nearly identical to the full-range RPA correlation energy. 

Turning now to the relativistic results, we observe a very different behavior. Namely, for $\tilde{\mu} \to \infty$, the full-range RRPA correlation energy per particle is linear with respect to $k_{\text{F}}$
\begin{eqnarray}
{\epsilon}_{\text{c}}^{\text{RRPA}}(n) \isEquivTo{k_{\text{F}} \to \infty} -0.0014 \; k_{\text{F}},
\label{eq:rel_full_range_high_density}
\end{eqnarray}
which is in agreement with other RRPA calculations reported in the literature~\cite{SchEngDreBlaSch-AQC-98,FacEngDreAndMul-PRA-98}. This is the ultra-relativistic limit, $\tilde{c} \to 0$, which is akin to a strong-correlation limit where both the exchange and correlation energies per particle are linear with respect to $k_{\text{F}}$. A similar linear behavior is also observed for the case of the long-range interaction. Again, for $\tilde{\mu} \gtrsim 20$ a.u., the long-range RRPA correlation energy is nearly identical to the full-range RRPA correlation energy.

\subsection{Relativistic long-range correlation factor}

We show in Fig.~\ref{fig:phi_lr} the relativistic long-range correlation factor ${\phi}_{\text{c}}^{\text{lr,RRPA},\tilde{\mu}}$ as a function of $k_{\text{F}}$ and $\tilde{\mu}$. We observe that, for all values of $\tilde{\mu}$ and all relevant values of $k_{\text{F}}$, the relativistic factor is greater than $1$, i.e. relativistic effects increase the magnitude of the correlation energy. Moreover, ${\phi}_{\text{c}}^{\text{lr,RRPA},\tilde{\mu}}$ is an increasing function of $k_{\text{F}}$, i.e. the relative relativistic effects increase as we increase the density.

In Fig.~\ref{fig:phi_lr} (a), it appears at first sight that ${\phi}^{\text{lr,RRPA},\tilde{\mu}}$ is a monotonic decreasing function of $\tilde{\mu}$, but the dependence on $\tilde{\mu}$ is in fact more complicated and is plotted in Fig.~\ref{fig:phi_lr} (b) for several values of $k_{\text{F}}$. For clarity, we show only values of $k_{\text{F}}$ lower than $200$ a.u., but the behavior is similar for the whole range of Fermi wave vectors that we have considered. For any value of $k_{\text{F}}$, it appears that ${\phi}^{\text{lr,RRPA},\tilde{\mu}}$ starts as an increasing function of $\tilde{\mu}$ until it reaches a maximum for a value $\tilde{\mu}^{\text{max}}(k_{\text{F}})$, after which it becomes a decreasing function of $\tilde{\mu}$ converging to its full-range interaction limit. The value of $\tilde{\mu}^{\text{max}}(k_{\text{F}})$ is itself an increasing function of $k_{\text{F}}$, going from $\tilde{\mu}^{\text{max}}(10) \approx 0.5$ to $\tilde{\mu}^{\text{max}}(1200) \approx 1.5$ a.u.. Furthermore, while ${\phi}^{\text{lr,RRPA},\tilde{\mu}}$ increases rapidly before $\tilde{\mu}^{\text{max}}(k_{\text{F}})$, it decreases only slightly afterward. This behavior explains why in Fig.~\ref{fig:phi_lr} (a) we observe that all curves for $\tilde{\mu}$ higher than $1$ appear to be superposed since there is little variation of ${\phi}^{\text{lr,RRPA},\tilde{\mu}}$ with respect to $\tilde{\mu}$ for these values, and why we observe a monotonic decreasing behavior with respect to $\tilde{\mu}$ only for lower values of $\tilde{\mu}$.
It appears that for $\tilde{\mu} \rightarrow 0$ the relativistic correction factor goes to $1$ for all values of $k_{\text{F}}$, i.e. the relativistic effects disappear when only the very long-range part of the electron-electron interaction remains. In this limit, however, the long-range correlation energy itself vanishes.

\section{Parametrization}

We now construct parametrizations of our numerical data. As building blocks for a parametrization of ${\phi}^{\text{lr,RRPA},\tilde{\mu}}$, we first parametrize the high-density limits of the non-relativistic and relativistic long-range correlation energies.

\subsection{High-density limit of the non-relativistic long-range correlation energy}

The parametrization of the high-density limit of the non-relativistic correlation energy is done by combining a parametrization for large values of $\tilde{\mu}$ and a parametrization for small values of $\tilde{\mu}$.

For sufficiently large values of $\tilde{\mu}$, the non-relativistic long-range RPA correlation energy per particle in the high-density limit follows a logarithmic behavior similar to the one of the non-relativistic full-range RPA correlation energy per particle [see Eq.~(\ref{eq:full_range_high_density})], and we found that the dependence on $\tilde{\mu}$ can be approximated by
	\begin{eqnarray}
{\epsilon}_{\text{c}}^{\text{lr,RPA},\tilde{\mu},\text{hd}_{1}}(n) = - \frac{1 - \ln 2}{{\pi}^{2}}\ln k_{\text{F}} \phantom{xxxxxxxxxxxxxx}
\nonumber\\
+ \left(-0.0508324 + \frac{1 + a_{1} \tilde{\mu}}{a_{2} + a_{3} \tilde{\mu} + a_{4} \tilde{\mu}^{2} + a_{5} \tilde{\mu}^{3}}\right),
	\label{eq:long_range_high_density}
\end{eqnarray}
giving our first high-density (hd$_1$) parametrization. The parameters $a_1=3.72862$, $a_2=3.53869$, $a_3=43.4382$, $a_4=40.2625$, and $a_5=53.1731$ have been fitted on numerical values of ${\epsilon}_{\text{c}}^{\text{lr,RPA},\tilde{\mu}}(n) + [(1 - \ln 2)/\pi^{2}]\ln k_{\text{F}}$ at $k_{\text{F}} = 9600~\text{a.u.}$ for 21 values of $\tilde{\mu} \geq 0.1$ a.u..

For sufficiently small values of $\tilde{\mu}$, the high-density limit of the non-relativistic RPA long-range correlation energy per particle can be approximated by the expression of Paziani \textit{et al.}~\cite{PazMorGorBac-PRB-06}
\begin{eqnarray}
	 {\epsilon}_{\text{c}}^{\text{lr,RPA},\tilde{\mu},\text{hd}_{2}}(n) 
	 = \frac{2\ln 2-2}{{\pi}^{2}}\ln\left[\frac{1 + b_{1}x + b_{2}x^{2} + b_{3}x^{3}}{1 + b_{1} x + b_{4}x^{2}} \right],~~  
\label{eq:long_range_high_density_paz}
\end{eqnarray}
with $x = {\mu}\sqrt{r_{\text{s}}} = (3\sqrt{\pi}/2)^{1/3} \tilde{\mu} \sqrt{k_{\text{F}}}$ and the parameters $b_1=5.84605$, $b_2=7.44953$, $b_3=3.91744$, and $b_4=3.44851$ are taken from Ref.~\onlinecite{PazMorGorBac-PRB-06}. This gives us our second high-density ($\text{hd}_{2}$) parametrization.

\begin{figure*}[t]
        \centering
         \includegraphics[width=5.9cm,angle=270]{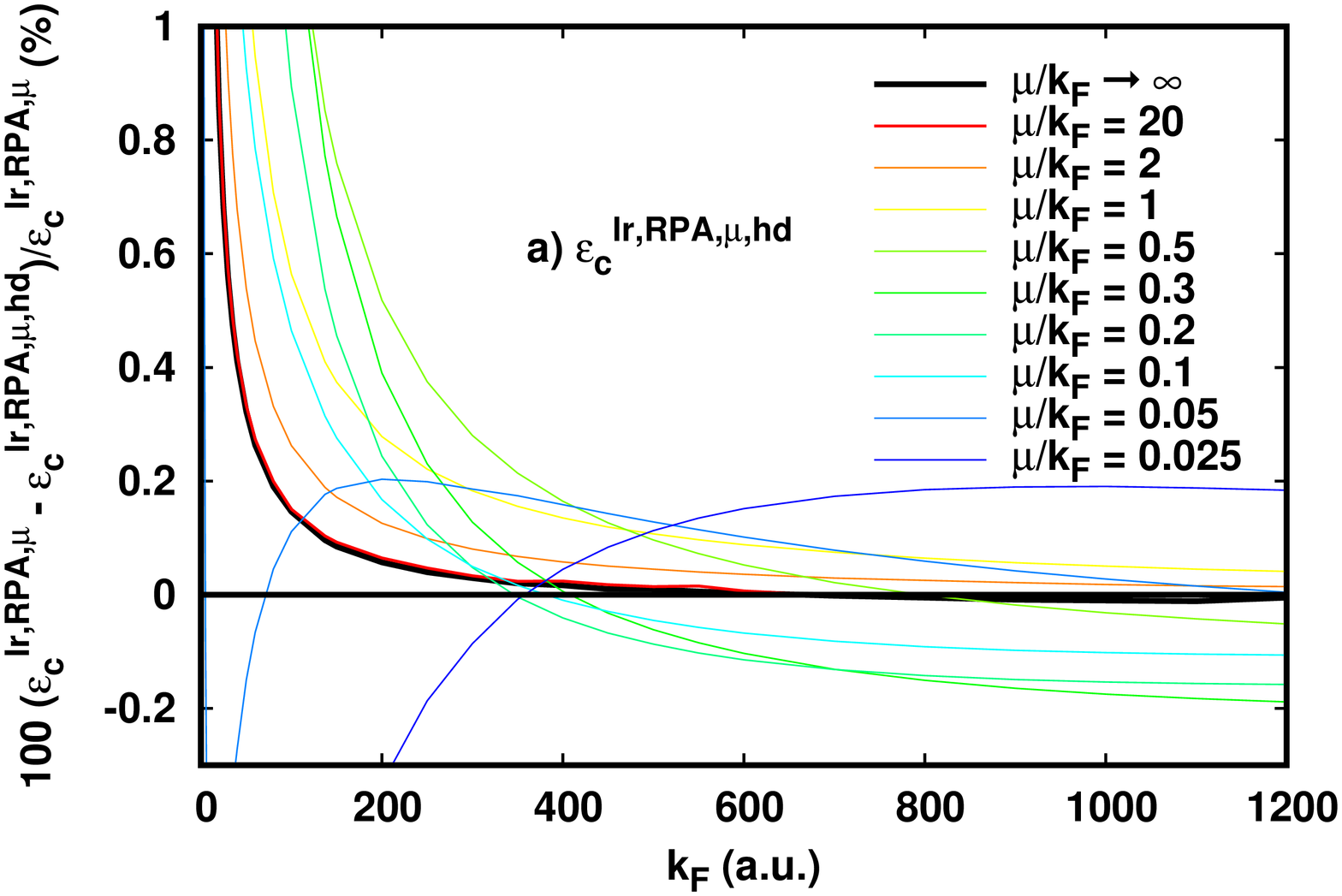}
        \includegraphics[width=5.9cm,angle=270]{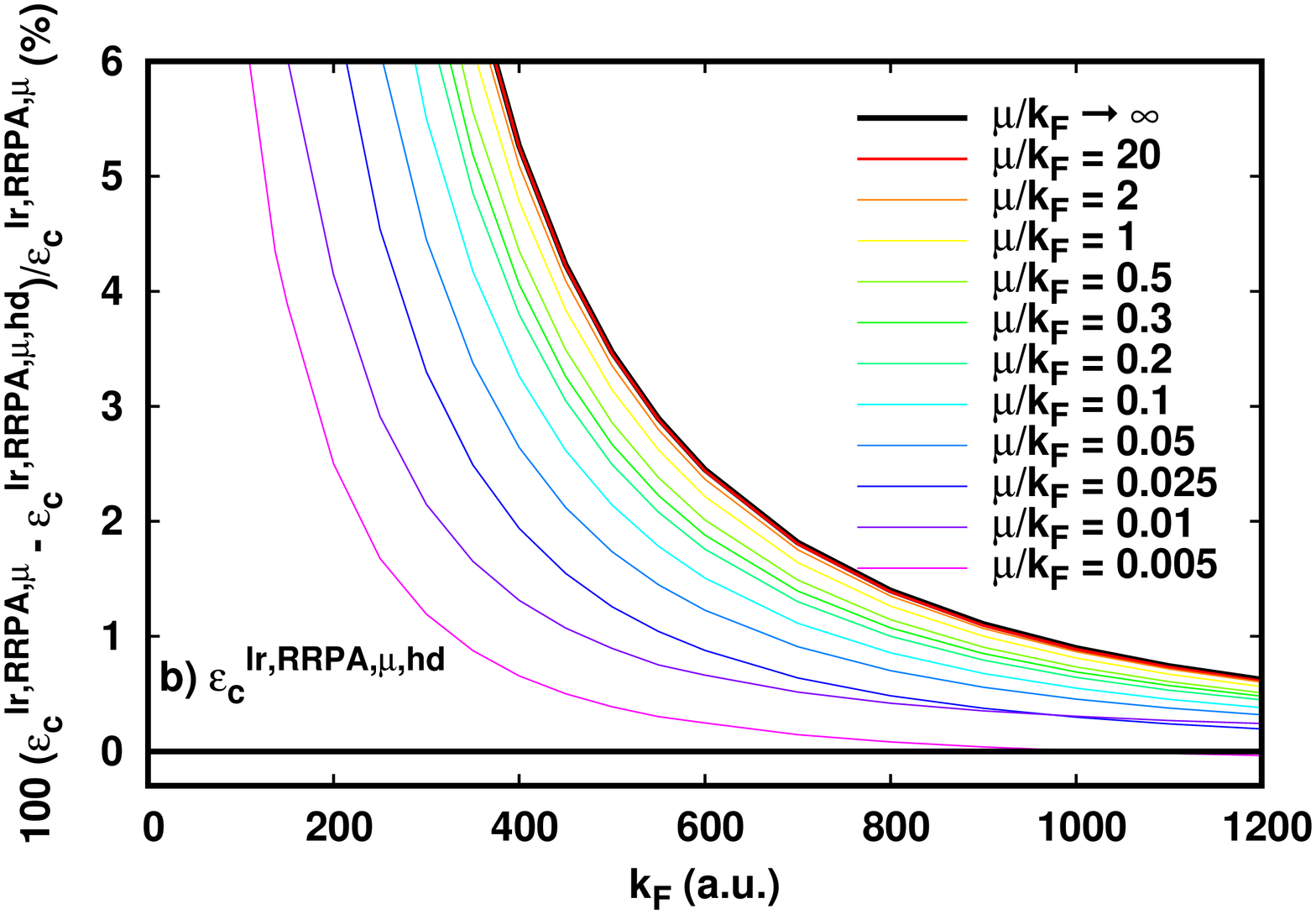}
\caption{Relative error of the high-density approximations for the non-relativistic (a) and relativistic (b) long-range RPA correlation energies per particle [Eqs.~(\ref{eq:long_range_high_density_continuation}) and~(\ref{eq:long_range_high_density_relativistic})].}
\label{fig:nr_hd_tier}
\end{figure*}

We now combine these two high-density parametrizations in a single parametrization by interpolating using the switching function $f(\tilde{\mu})=\erf(3\tilde{\mu})^{4}$
\begin{eqnarray}
{\epsilon}_{\text{c}}^{\text{lr,RPA},\tilde{\mu},\text{hd}}(n) &=& f(\tilde{\mu}) {\epsilon}_{\text{c}}^{\text{lr,RPA},\tilde{\mu},\text{hd}_{1}}(n) 
\nonumber\\
&&
+ (1 - f(\tilde{\mu}) ){\epsilon}_{\text{c}}^{\text{lr,RPA},\tilde{\mu},\text{hd}_{2}}(n). ~~~~
	\label{eq:long_range_high_density_continuation}
\end{eqnarray}
The use of the fourth power of the error function allows for a steep enough switching and using $3\tilde{\mu}$ as the argument puts the transition between the two parts around $\tilde{\mu} = 0.3$ a.u.. Equation~(\ref{eq:long_range_high_density_continuation}) thus constitutes our high-density approximation for the non-relativistic long-range RPA correlation energy per particle valid for all $\tilde{\mu}$. In particular, for $\tilde{\mu}\to\infty$, it correctly reduces to the full-range behavior in Eq.~(\ref{eq:full_range_high_density}).

The relative error of this high-density approximation ${\epsilon}_{\text{c}}^{\text{lr,RPA},\tilde{\mu},\text{hd}}(n)$ is plotted in Fig.~\ref{fig:nr_hd_tier} (a). For $k_\text{F} \gtrsim 400$ a.u. and $\tilde{\mu} \geq 0.025$ a.u., the high-density approximation gives a relative error of less than $0.2\%$. For smaller values of $\tilde{\mu}$ (not shown), the maximal relative error increases up to around $3\%$ but the error is made on very small values of the correlation energy.

\begin{figure}[t]
        \centering
         \includegraphics[width=5.9cm,angle=270]{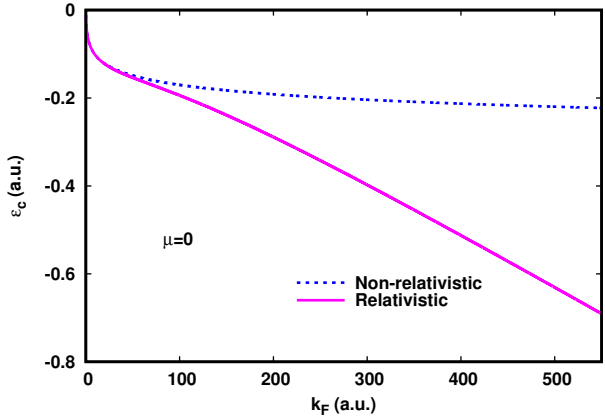}
	\caption{Non-relativistic and relativistic complementary short-range correlation energies per particle of the HEG.}
\label{fig:ec}
\end{figure}

\subsection{High-density limit of the relativistic long-range correlation energy}

In the high-density limit, the relativistic long-range RRPA correlation energy is linear in $k_{\text{F}}$ for all values of $\tilde{\mu}$ and it is well approximated by
\begin{eqnarray}
 {\epsilon}_{\text{c}}^{\text{lr,RRPA},\tilde{\mu},\text{hd}}(n) = -0.185345 \phantom{xxxxxxxxxxxxxxxxxxxx}
\nonumber\\
\times\left(1 - \frac{1 + c_{1} \tilde{\mu} + c_{2}\tilde{\mu}^{2} + c_{3}\tilde{\mu}^{3} + c_{4}\tilde{\mu}^{4}}{1 + c_{5} \tilde{\mu} + c_{6} \tilde{\mu}^{2} + c_{7} \tilde{\mu}^{3} + c_{8} \tilde{\mu}^{4} + c_{9}\tilde{\mu}^{5}}\right)/\tilde{c},
	\label{eq:long_range_high_density_relativistic}
\end{eqnarray}
where the parameters $c_1=63.6213$, $c_2=161.703$, $c_3=58.4589$, $c_4=-0.55375$, $c_5=63.7034$, $c_6=467.578$, $c_7=624.653$, $c_8=952.370$, and $c_9=159.956$ have been obtained by fitting at $k_{\text{F}} = 9600~\text{a.u.}$ using all 26 values for $\tilde{\mu}$ considered in this work. For $\tilde{\mu} \to \infty$, Eq.~(\ref{eq:long_range_high_density_relativistic}) correctly reduces to the full-range behavior in Eq.~(\ref{eq:rel_full_range_high_density}).

The relative error of this high-density approximation ${\epsilon}_{\text{c}}^{\text{lr,RRPA},\tilde{\mu},\text{hd}}(n)$ is plotted in Fig.~\ref{fig:nr_hd_tier} (b). For $\tilde{\mu} \to \infty$, the relative error gets below 1\% for $k_\text{F} \gtrsim 1000$ a.u. As $\tilde{\mu}$ decreases, the high-density regime is reached for smaller values of $k_{\text{F}}$, e.g. for $\tilde{\mu} = 0.005$ a.u. we obtain 1\% accuracy for $k_\text{F} \gtrsim 300$ a.u.

\subsection{Parametrization of the relativistic long-range correlation factor}

 \begin{table*}[t]
 \caption{Parameters for the relativistic long-range correlation factor ${\phi}_{\text{c}}^{\text{lr,RRPA},\tilde{\mu}}$ [Eq.~(\ref{eq:fit_philr})].}
  \centering
     \begin{tabular}{lllllll}
  \hline\hline
        	$i$ & $~~~~a_{1,i}$ & $~~~~a_{2,i}$ & $~~~~a_{3,i}$ & $~~~~b_{1,i}$ & $~~~~b_{2,i}$ & $~~~~b_{3,i}$  \\
  \hline 
  1 & 2.22080~$\times10^{-2}$ & 9.66045~$\times10^{-2}$ & 1.59065~$\times10^{-4}$  & ~~~~~ -     & ~~~~~ -     &  ~~~~~ -     \\
  2 & 7.04721~$\times10^{-1}$ & 2.66457     & 9.62993~$\times10^{-2}$  & 7.09439~$\times10^{-1}$ & 2.91597~$\times10^{-1}$ & -2.40333~$\times10^{-3}$ \\
  3 & ~~~~ -      & 9.24891~$\times10^{-1}$ & 6.30881~$\times10^{-1}$  & ~~~~~ -     & 5.62594~$\times10^{-1}$ &  6.077222~$\times10^{-3}$\\
  4 & 1.16165~$\times10^{-1}$ & 1.50127~$\times10^{-1}$ & 5.30353~$\times10^{-3}$  & ~~~~~ -     & ~~~~~ -     & ~~~~~ -      \\
  5 & ~~~~~ -     & 3.07852     & 5.32685~$\times10^{-1}$  & ~~~~~ -     & 7.56679~$\times10^{-1}$ &  8.30363~$\times10^{-1}$ \\
  \hline\hline
 \end{tabular}
 \label{table:phi_lr_fit}
 \end{table*}

\begin{figure}[t]
        \centering
         \includegraphics[width=5.9cm,angle=270]{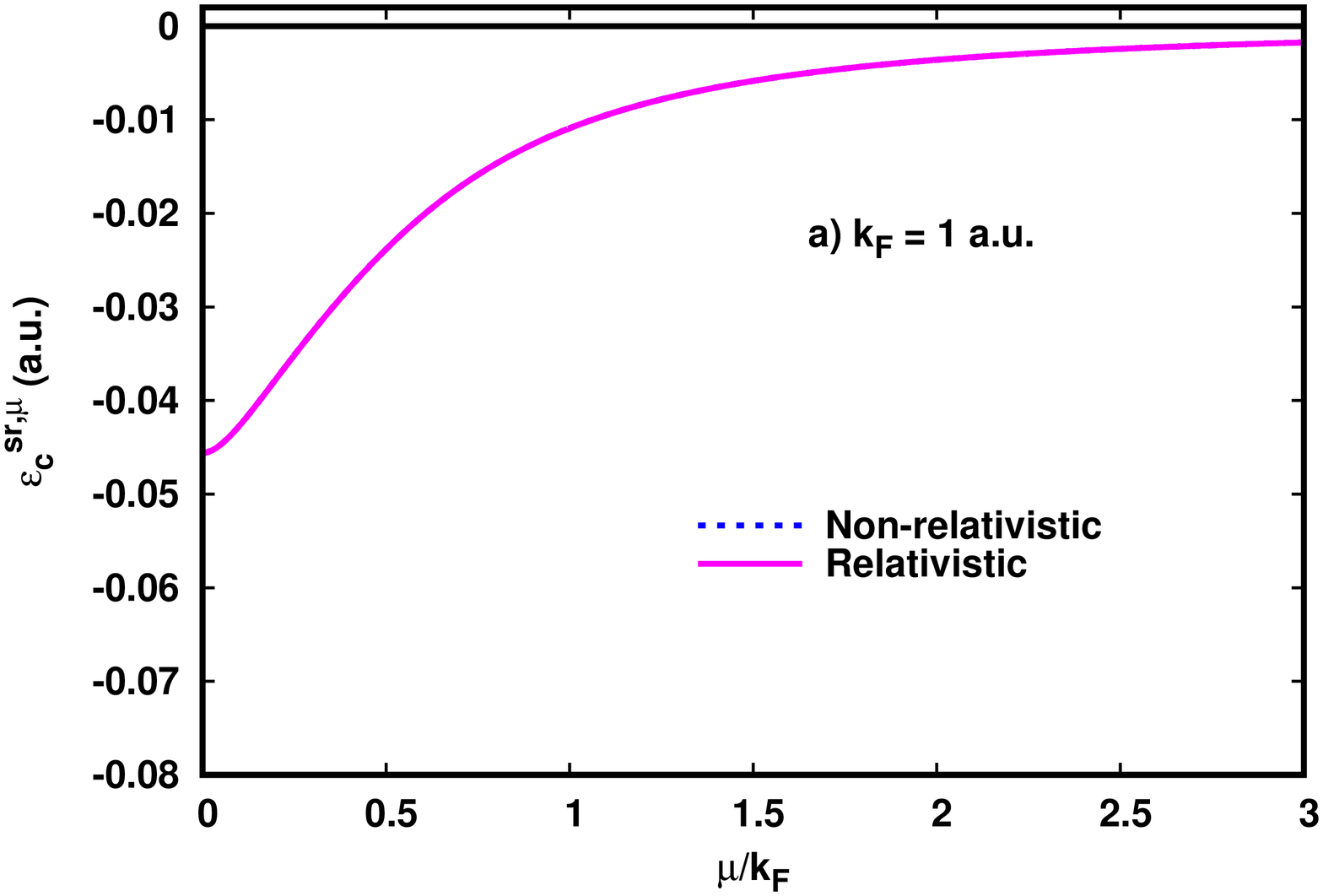}
         \includegraphics[width=5.9cm,angle=270]{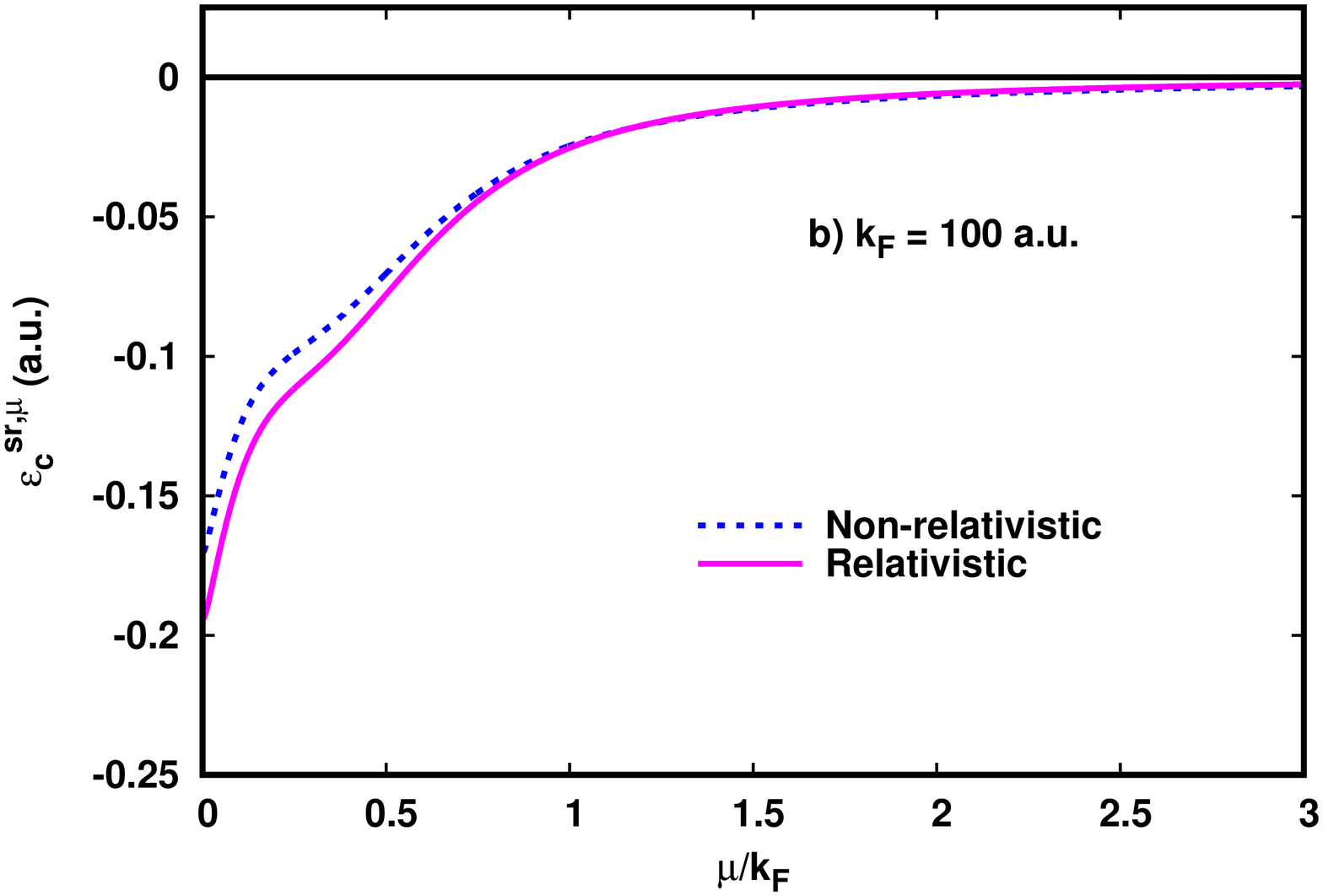}
         \includegraphics[width=5.9cm,angle=270]{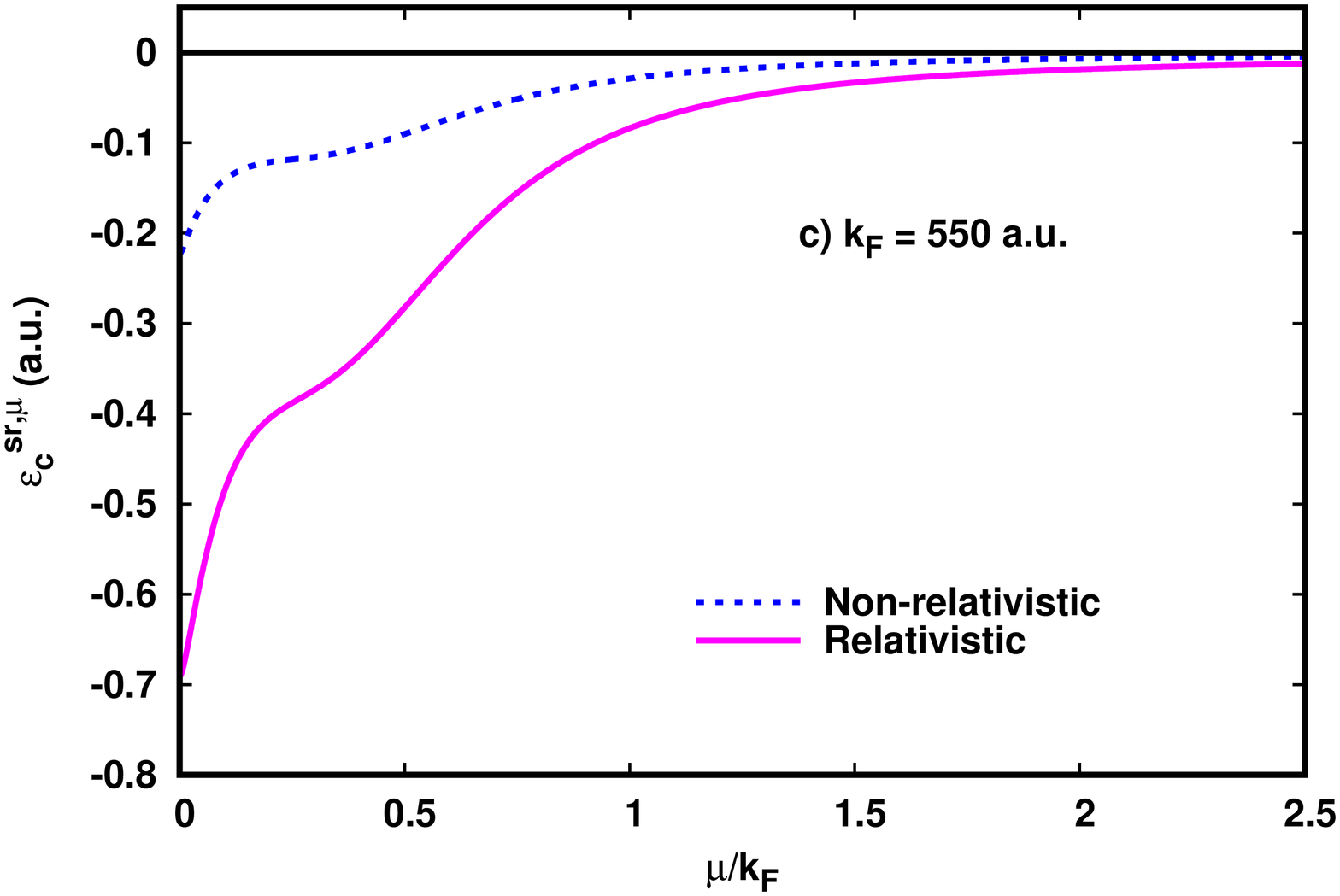}
\caption{Non-relativistic and relativistic complementary short-range correlation energy per particle of the HEG for $k_\text{F}=1$ a.u. (a), $k_\text{F}=100$ a.u. (b), and $k_\text{F}=550$ a.u. (c).}
\label{fig:sr_ec}
\end{figure}

Having found parametrizations for the high-density limit of the non-relativistic and relativistic long-range RPA correlation energies per particle, we now use these expressions to build a Pad\'e-like expression for the relativistic long-range correlation factor ${\phi}_{\text{c}}^{\text{lr,RRPA},\tilde{\mu}}$. We found that it is accurately represented by
\begin{widetext}
\begin{eqnarray}
{\phi}_{\text{c}}^{\text{lr,RRPA},\tilde{\mu}}(n) = \frac{1 + \displaystyle{\frac{a_{1,1} + a_{1,2}\tilde{\mu}}{a_{1,4} + \tilde{\mu}}}/\tilde{c} + \frac{a_{2,1} + a_{2,2}\tilde{\mu} + a_{2,3}\tilde{\mu}^{2}}{a_{2,4} + a_{2,5}\tilde{\mu} + \tilde{\mu}^{2}}/\tilde{c}^{2} + \frac{a_{3,1} + a_{3,2}\tilde{\mu} + a_{3,3}\tilde{\mu}^{2}}{a_{3,4} + a_{3,5}\tilde{\mu} + \tilde{\mu}^{2}}/\tilde{c}^{3} - {\epsilon}_{\text{c}}^{\text{lr,RRPA},\tilde{\mu},\text{hd}}(n) /\tilde{c}^{4}}{1 + \displaystyle{\frac{a_{1,1} + b_{1,2}\tilde{\mu}}{a_{1,4} + \tilde{\mu}}}/\tilde{c} + \frac{a_{2,1} + b_{2,2}\tilde{\mu} + b_{2,3}\tilde{\mu}^{2}}{a_{2,4} + b_{2,5}\tilde{\mu} + \tilde{\mu}^{2}}/\tilde{c}^{2} + \frac{a_{3,1} + b_{3,2}\tilde{\mu} + b_{3,3}\tilde{\mu}^{2}}{a_{3,4} + b_{3,5}\tilde{\mu} + \tilde{\mu}^{2}}/\tilde{c}^{3} - {\epsilon}_{\text{c}}^{\text{lr,RPA},\tilde{\mu},\text{hd}}(n)/\tilde{c}^{4}}. 
\label{eq:fit_philr}
\end{eqnarray}
\end{widetext}
The choice of using the opposite of the high-density correlation energies as coefficients of $1/\tilde{c}^4$ terms ensures that these coefficients are positive and reduces the risk of introducing poles within the parametrization. The parameters are given in Table~\ref{table:phi_lr_fit}. They have been found by fitting to the numerical values of ${\epsilon}_{\text{c}}^{\text{lr,RRPA},\tilde{\mu}}/{\epsilon}_{\text{c}}^{\text{lr,RPA},\tilde{\mu}}$ using all values of $k_\text{F}$ and $\tilde{\mu}$ considered in this work. The maximal absolute error is less than $0.4\%$ for the smallest values of $\tilde{\mu}$ considered. In the special case $\tilde{\mu} \to \infty$, we obtain the full-range relativistic correlation factor ${\phi}_{\text{c}}^{\text{lr,RRPA},\tilde{\mu}\to\infty}(n) = {\phi}_{\text{c}}^{\text{RRPA}}(n)$, with a maximal absolute error less than $0.1\%$. Again, we stress that the parametrization of Eq.~(\ref{eq:fit_philr}) is valid for the physical value of the speed of light $c$, and not for an arbitrary value of $c$. For more details on the fit, see Ref.~\onlinecite{Paq-THESIS-20}.

 \subsection{Complementary short-range correlation energy per particle}

From Eqs.~(\ref{epscsrRHEG})-(\ref{eq:long_range_correlation_density_functional}), we finally obtain our approximation for the complementary short-range correlation energy per particle of the RHEG 
\begin{eqnarray}
\bar{\epsilon}_{\text{c}}^{\text{sr,RHEG},{\mu}}(n) 
                        &\approx& {\epsilon}_{\text{c}}^{\text{HEG}}( n) {\phi}_{\text{c}}^\text{RRPA}(n) - {\epsilon}_{\text{c}}^{\text{lr,HEG},{\mu}}(n) {\phi}_{\text{c}}^{\text{lr},\tilde{\mu},\text{RRPA}}(n),
\nonumber\\
\label{eq:sr_ec}
\end{eqnarray}
in which we use the Perdew-Wang-92 parametrization for ${\epsilon}_{\text{c}}^{\text{HEG}}(n)$~\cite{PerWan2-PRB-92} and the parametrization of Paziani \textit{et al.}~\cite{PazMorGorBac-PRB-06} for ${\epsilon}_{\text{c}}^{\text{lr,HEG},{\mu}}(n)$.

In the limit $\mu=0$, this short-range correlation energy per particle reduces to the full-range correlation energy per particle, i.e. $\bar{\epsilon}_{\text{c}}^{\text{sr,RHEG},{\mu}=0}(n) = \epsilon_{\text{c}}^{\text{RHEG}}(n)$. In Fig.~\ref{fig:ec}, we compare our obtained $\epsilon_{\text{c}}^{\text{RHEG}}(n)$ with its non-relativistic analog ${\epsilon}_{\text{c}}^{\text{HEG}}(n)$. As already indicated, relativistic effects increase the magnitude of the correlation energy for large densities and turn the logarithmic dependence with respect to $k_\text{F}$ into a linear dependence.

We plot in Fig.~\ref{fig:sr_ec} the relativistic and non-relativistic complementary short-range correlation energies per particle as a function of $\tilde{\mu}$ for several values of $k_\text{F}$. For $k_\text{F}=100$ a.u., we already see the impact of the relativistic effects for small values of $\tilde{\mu}$. For $k_\text{F}=550$ a.u., the relativistic effects are important for all relevant values of $\tilde{\mu}$. Note that the wiggling behavior with respect to $\tilde{\mu}$ observed on the graphs for $k_\text{F}=100$ and 550 a.u. is most likely unphysical and comes from the parametrization of the non-relativistic long-range correlation energy per particle ${\epsilon}_{\text{c}}^{\text{lr,HEG},{\mu}}(n)$. This is not so surprising since such high densities were not considered in the construction of the parametrization of Ref.~\onlinecite{PazMorGorBac-PRB-06}. This calls perhaps for a refinement of this parametrization. For high enough densities, however, the possible refinement of ${\epsilon}_{\text{c}}^{\text{lr,HEG},{\mu}}(n)$ is secondary in comparison to the relativistic effects.

Finally, we mention another possible limitation of our parametrization: we did not impose the large-$\mu$ behavior of the complementary short-range correlation energy per particle of the RHEG, which is expected to have the same form as its non-relativistic analog~\cite{TouColSav-PRA-04,GorSav-PRA-06,PazMorGorBac-PRB-06} (as the large-$\mu$ behavior of the relativistic and non-relativistic short-range exchange energies had the same form~\cite{PaqGinTou-JCP-20}), i.e.
\begin{eqnarray}
\bar{\epsilon}_{\text{c}}^{\text{sr,RHEG},{\mu}}(n) \isEquivTo{\mu \to \infty} \frac{k_\text{F}^3 \; g_\text{c}^{\text{RHEG}}(0,n)}{6\pi \mu^2},
\label{}
\end{eqnarray}
where $g_\text{c}^{\text{RHEG}}(0,n)$ is the correlation contribution to the on-top pair-distribution function of the RHEG. Indeed, we do not have a good estimate of $g_\text{c}^{\text{RHEG}}(0,n)$ and RRPA is not expected to be accurate for this quantity. Therefore, we do not expect our parametrization to be very accurate for large $\mu$. Fortunately, the short-range correlation energy is small anyway for large $\mu$.

\section{Conclusion}

From RRPA calculations on the RHEG, we have constructed the complementary short-range correlation RLDA functional to be used in relativistic RS-DFT based on a Dirac-Coulomb Hamiltonian in the no-pair approximation. This short-range correlation RLDA functional could be tested on atomic and molecular systems, and will most likely serve a starting point for building more sophisticated relativistic short-range correlation functionals, e.g. depending on the density gradient or on the on-top pair density as already done for the short-range exchange functional~\cite{PaqGinTou-JCP-20}. We believe that the present work helps to establish relativistic RS-DFT on a firm ground and will eventually be useful for electronic-structure calculations of strongly correlated systems containing heavy elements.

\section*{Data Availability Statement}
Data available on request from the authors.

\appendix

\section{Non-interacting linear-response function of the RHEG in the no-pair approximation}
\label{app:linear_response_function}

The non-interacting one-electron Green function of the RHEG in the no-pair approximation at wave vector $\b{k}$ and frequency $\omega$ is
\begin{eqnarray}
G_{0}(\b{k},{\omega}) &=&  \sum_{s\in\{\uparrow,\downarrow\}} {\psi}_{\b{k},{s}}^{} {\psi}_{\b{k},{s}}^{\dagger} \left[\frac{{\theta}(k-k_{\text{F}})}{{\omega} - {\varepsilon}_{k} + i0^+} + \frac{{\theta}(k_{\text{F}} - k)}{{\omega} - {\varepsilon}_{k} - i0^+}\right],
\nonumber\\
\end{eqnarray}
where $k=|\b{k}|$ and ${\psi}_{\b{k},{s}}$ are the four-component spinors associated with the positive-energy solutions of the non-interacting Dirac equation
\begin{eqnarray}
{\psi}_{\b{k},{s}} = \sqrt{\frac{\varepsilon_{k}+c^{2}}{2\varepsilon_{k}}} \left(\begin{array}{c}
                {\varphi}_{s}\\
		\frac{c (\bm{\sigma} \cdot \b{k})}{\varepsilon_{k} + c^{2}}{\varphi}_{s}
\end{array}\right),
\end{eqnarray}
where $\bm{\sigma}$ is the vector composed of the three Pauli matrices, $\varepsilon_{k} = \sqrt{k^{2}c^{2} + c^{4}}$ are the one-electron energies, and ${\varphi}_{s}$ are the two-component spinors
\begin{eqnarray}
{\varphi}_{\uparrow} = \left( \begin{array}{c} 1\\0\end{array} \right) \;\; \text{and} \;\; {\varphi}_{\downarrow} = \left( \begin{array}{c} 0\\1\end{array} \right).
\end{eqnarray}
Note that the Green function $G_{0}(\b{k},{\omega})$ is a 4$\times$4 matrix.

The corresponding no-pair longitudinal (i.e. density-density) non-interacting linear-response function at wave vector $q=|\b{q}|$ and frequency $q_0$ is
\begin{eqnarray}
{\chi}_{0}(q,q_{0})
	&=& \int \! \frac{\text{d}\b{k}}{(2{\pi})^{3}} \int_{-\infty}^{+\infty} \! \frac{\text{d}{\omega}}{2\pi i} ~\text{Tr} \left[G_{0}(\b{k},{\omega})G_{0}(\b{k}+\b{q},{\omega}+q_{0})\right],
\nonumber\\
\end{eqnarray}
which, after calculating the trace of the products of spinors (see, e.g., Ref.~\onlinecite{PaqTou-JCP-18}) and calculating the integral over $\omega$ by contour integration, gives
\begin{widetext}
\begin{eqnarray}
{\chi}_{0}(q,q_{0}) &=& \int \frac{\text{d}\b{k}}{(2{\pi})^{3}} \Bigg( 1 + \frac{\b{k} \cdot(\b{k+q})c^{2} + c^{4}}{\varepsilon_{k} \varepsilon_{|\b{k}+\b{q}|}} \Bigg){\theta}(|\b{k+q}|-k_{\text{F}}){\theta}(k_{\text{F}} - k)
        \Bigg[ \frac{-1}{q_{0} + {\varepsilon}_{|\b{k+q}|} - {\varepsilon}_{k} - i0^+} + \frac{1}{q_{0} + {\varepsilon}_{k} - {\varepsilon}_{|\b{k+q}|} + i0^+}  \Bigg].
\end{eqnarray}
Evaluating the linear-response function at imaginary frequency $q_0=iu$, and after simplifying, we find
\begin{eqnarray}
{\chi}_{0}(q,iu) &=& - \int \frac{\text{d}\b{k}}{(2{\pi})^{3}} {\theta}(k_{\text{F}}-k) \Bigg(1+ \frac{\b{k}\cdot(\b{k+q})c^{2} + c^{4}}{\varepsilon_{k}\varepsilon_{|\b{k}+\b{q}|}} \Bigg) \frac{2(\varepsilon_{|\b{k+q}|} - \varepsilon_{k})}{u^{2}+( \varepsilon_{|\b{k+q}|} - \varepsilon_{k})^{2}},
\end{eqnarray}
which can also be written as
\begin{eqnarray}
{\chi}_{0}(q,iu) &=&  - \int \frac{\text{d}\b{k}}{(2{\pi})^{3}} {\theta}(k_{\text{F}}-k) \frac{\left[ \big(\varepsilon_{k} + \varepsilon_{|\b{k}+\b{q}|} \big)^{2} - q^{2}c^{2}\right] \big(\varepsilon_{|\b{k+q}|} - \varepsilon_{k}\big)}{\varepsilon_{k}\varepsilon_{|\b{k}+\b{q}|} \left[ u^{2}+( \varepsilon_{|\b{k+q}|} - \varepsilon_{k})^{2}\right]}.
	\label{eq:chi_0}
\end{eqnarray}
\end{widetext}
This expression is equal, up to a trivial sign convention, to the first term of the longitudinal non-interacting linear-response function given by Ramana and Rajagopal~\cite{RamRaj-PRA-81} [Eq.~(6) of Ref.~\onlinecite{RamRaj-PRA-81}]. The expression determined in their work is not within the no-pair approximation but within the no-sea approximation, and thus their expression includes a renormalization term coming from the negative-energy states. The no-pair longitudinal non-interacting linear-response function of the RHEG was also calculated by Facco Bonetti \textit{et al.}~\cite{FacEngDreAndMul-PRA-98}, who gave a closed-form expression for real frequencies [Eq.~(A1) of Ref.~\onlinecite{FacEngDreAndMul-PRA-98}]. However, to the best of our knowledge, their expression cannot be straightforwardly used for imaginary frequencies. We prefer then to use Eq.~(\ref{eq:chi_0}) in order to work with imaginary frequencies. After introducing adimensional variables and simplifying, Eq.~(\ref{eq:chi_0}) leads to Eq.~(\ref{eq:linear_response_fct}) and we perform the integral numerically. For more details on the derivation of Eq.~(\ref{eq:chi_0}), see Ref.~\onlinecite{Paq-THESIS-20}.

\end{document}